# Topological and Network Analysis of Lithium Ion Battery Components: The Importance of Pore Space Connectivity for Cell Operation

M. F. Lagadec,[1] R. Zahn,[1] S. Müller[1] and V. Wood[1]

[1] Department of Information Technology and Electrical Engineering, ETH Zurich, Zurich CH-8092, Switzerland.

The structure of lithium ion battery components, such as electrodes and separators, are commonly characterised in terms of their porosity and tortuosity. The ratio of these values gives the effective transport of lithium ions in the electrolyte-filled pore spaces, which can be used to determine the ionic resistivity and corresponding voltage losses. Here, we show that these microstructural characteristics are not sufficient. Analysis of tomographic data of commercial separators reveals that different polyolefin separators have similar porosity and through-plane tortuosity, which, in the homogenised picture of lithium ion cell operation, would imply that these different separators exhibit similar performance. However, numerical diffusion simulations indicate that this is not the case. We demonstrate that the extent to which lithium ion concentration gradients are induced or smoothed by the separator structure is linked to pore space connectivity, a parameter that can be determined by topological or network based analysis of separators. These findings enable us to propose how to design separator microstructures that are safer and accommodate fast charge and discharge.

The structures of components in a lithium ion battery (LIB), such as the electrodes and the separator, influence lithium ion transport[1] and therefore play an important role in dictating the cell performance metrics such as (dis)charge-rate dependent capacity and cycle life[2].

In the homogenised picture of cell operation used in 1D models[3–5] that dominate cell modelling today (e.g., Dualfoil[6] and COMSOL Multiphysics[7]), the diffusion coefficient of the cations ($D_+$) and the anions ($D_-$) in the electrolyte-filled pore space is given by their diffusion in a bath of electrolyte scaled by the effective transport coefficient of the microstructure,[8] $\delta_{TP} = \varepsilon / \tau_{TP}$, where $\varepsilon$ is the porosity and $\tau_{TP}$ is the tortuosity along the through-plane (TP) direction between the current collectors. A low effective transport coefficient leads to a low ionic diffusivity and therefore a low ionic conductivity ($\sigma \approx c \cdot (D_+ + D_-)$, where $c$ is the concentration of the salt in the electrolyte), which in turn results in large voltage drops (i.e., large overpotentials) across the electrolyte-filled pore space[9]. At fast operation speeds (e.g., at the end of a 5C discharge),[10] these overpotentials can account for ~60 % of cell overpotentials, outweighing the contributions of the charge transference resistance at the electrodes.

However, this volume-averaged effective transport in the homogenised picture does not account for inhomogeneities across the cells. Inhomogeneities lead to incomplete capacity extraction, lithium plating, and hot spots where current preferentially flows.[11–14] While inhomogeneities can be determined by running simulations over real 3D microstructures[8,11] or statistically assessing many subvolumes of an imaged microstructure[15], these analyses do not tell us about how the structure itself may give rise to or how good it is at compensating for inhomogeneities.

In this article, we propose a new approach to characterise microstructure of lithium ion battery components based on topological and network analysis. We show that this analysis captures how a structure induces or homogenises ion gradients.

While topological analysis of porous media is commonly used in soil physics and geology[16,17], it has not previously been applied to the LIB field. Linked to topology is network theory, which describes the types of connections that exist in a system that can be characterised by branches and nodes (i.e., points where branches intersect). Network analysis has been applied in a wide variety of fields including information and communication (e.g., the world-wide web), energy (e.g., power grids), and biology (e.g., metabolic networks),[18] but not to describe the pore space of a battery, which can also be viewed as a network.

Here, we show that parameters that can be calculated from topological and network analysis of 3D microstructures, such as pore space connectivity density and percent of dead end pores, are important for predicting cell performance and safety.

As a case study, we look at lithium ion battery separators. We show that two separators of strikingly different morphology have similar TP effective transport coefficients, suggesting that both separators would exhibit similar lithium ion transport. However, 3D diffusion simulations highlight that lithium ion transport occurs in different ways in the separator structures. We show that the differences in pore space topology and network properties of the two separators can explain the different transport properties in the separators, particularly the tendency of a structure to allow or prevent lithium ion concentration gradients. These parameters can be used to optimise separator selection for a given cell and to guide design of next generation separators.

Microporous polyolefin membranes have been used as separators in LIBs for several decades, and have been manufactured with a variety of thicknesses, pore structures, and surface chemistries.[19,20] Recently, we have shown that it is possible to obtain quantitative reconstructions of LIB separators using focus-ion-beam scanning electron microscope (FIB-SEM) tomography.[15,21] 3D microstructure renderings of polyethylene (PE)[22] and polypropylene (PP)[23] separators obtained using this approach are shown in **Figure 1**.

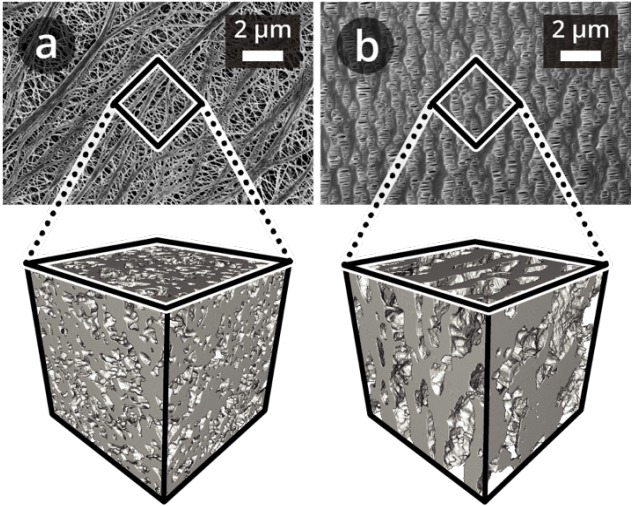

**Figure 1.** SEM top view image recorded in secondary electron mode, and 3D microstructure renderings of (a) Targray PE16A and (b) Celgard® PP1615 separators of sub-volumes of 3 μm edge length, imaged as described by Lagadec et al.[15]

The PE and PP separators exhibit distinct morphologies that stem from the different processes used to manufacture them. The PE separator (**Figure 1a**) microstructure is isotropic,[15] while that of PP is anisotropic[21] (**Figure 1b**). However, the respective porosities, ε, TP tortuosities, $\tau_{TP}$, and thus the effective transport coefficients[8], $\delta_{TP} = \varepsilon / \tau_{TP}$, of the PE and PP microstructures are similar (**Table I**).

Topological and network analysis of these structures provides a set of parameters with which to quantify separators. Here, we provide a brief introduction of these parameters for readers unfamiliar with morphological or network descriptors and illustrate simple cases in **Figure 2**.

The topological invariant (Euler-Poincaré characteristic, see sections 2-3 in the **ESI**)[24], X, describes an object's shape and structure independent of how it is bent and relates to the object's connectivity,[16,25] which is a concept from topology and network theory. The skeleton of the structure (i.e., pink lines in **Figure 2**) can be used to analyse the separator as a network.

X of a given pore network is N-C[†], where N is the number of unique pores, and C is their connectivity, which is defined as the number of cuts needed to obtain a simply connected network (i.e., without redundant connections).[26] From network analysis, C is also defined as the number of branches minus the end point branches (i.e., dead-end branches connected only to a single node) minus number of nodes plus 1.

To illustrate this connection between the topological invariant, X, and the connectivity, C, we consider two cases. In **Figure 2a**, we have N unconnected pores. C is zero and X is positive (X = N). In the case of a single pore network in **Figure 2b**, N = 1 and C = 2 (the two redundant connections are marked with cyan cuts) such that X is negative (X = 1 - 2 = -1). Alternatively, we see that there are 11 branches, 5 end points, and 5 nodes, also giving C = 2. A more connected network (higher C) implies a more negative X (**Figure 2c**).

From network analysis, we additionally consider the node density, the number of nodes of different order (the order is given by the number of branches connected to the node), the number of end point branches, and the average branch length.

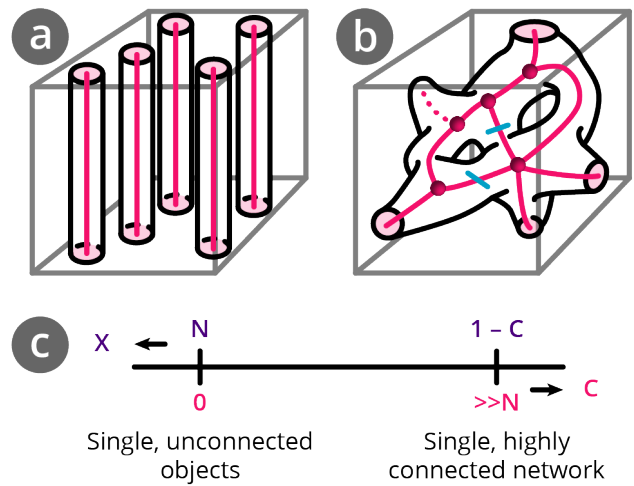

**Figure 2.** Schematics of pore space and pore space skeletonisation of (a) unconnected single objects with connectivity C = 0, and of (b) single, interconnected network (redrawn from DeHoff et al.[27] with branches and nodes) connectivity C = 2. (c) Relationship between Euler-Poincaré characteristic, X, and connectivity, C.

**Table I**. Porosity, ε, tortuosity, τ, and effective transport coefficient, δ, for representative volume elements of Targray PE16A [22] (PE) of 2 μm edge length and of Celgard® PP1615 [23] (PP) of 3 μm edge length. The tortuosity values are obtained from Fickian diffusion simulations across the pore phase in both in-plane directions (IP1 and IP2) and in the TP direction.

| Parameter | | PE | PP |
|---|---|---|---|
| Porosity ε [%] | | 40.82±1.92 | 40.19±1.03 |
| Tortuosity τ [-] | $\tau_{IP1}$ | 2.99±0.39 | 2.31±0.24 |
| | $\tau_{IP2}$ | 2.65±0.31 | 24.89±6.15 |
| | $\tau_{TP}$ | 2.64±0.21 | 2.04±0.19 |
| Effective transport coefficient δ [%] | $\delta_{IP1}$ | 13.9±2.2 | 17.6±2.1 |
| | $\delta_{IP2}$ | 15.7±2.2 | 1.7±0.4 |
| | $\delta_{TP}$ | 15.6±1.9 | 19.9±2.0 |

In the example in **Figure 2b**, we have one node of order 5 and four nodes of order 3. There are 5 end point branches, but, for our purposes, because we only work with a sub-volume of a separator, we count only those that end within the structure as end-point branches (dashed line).

Since the real separator structures are complex, we work with computer-generated, idealised structures as well as with the imaged PE and PP structures.

We generate structures (**Figure 3**) with cylindrical pores in 1, 2, and 3 directions using the algorithm described in section 4 of the **ESI**. For each type of structure, 3 entities are generated and the reported properties are the average values. The pore (i.e., cylinder) diameter is comparable to the geometrical pore size, $D_{50}$, of the PE separator, and their porosity, $\varepsilon$, is set to be within 40±2 %, which is comparable to the porosity of the PE and PP separators. These cubic datasets have an edge length of 5 μm and an isotropic voxel length of 10 nm.

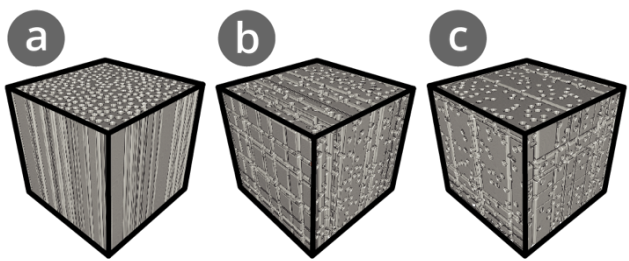

**Figure 3.** Example 3D microstructure renderings of artificial separator microstructures with randomly distributed, cylindrical pores of diameter 0.13 μm and edge length 3 μm in (a) one, (b) two, and (c) three dimensions.

For the three computer generated reference structures as well as the imaged PE and PP, we calculate X using the MATLAB code by Legland et al.[25] For the skeleton analysis, the datasets are symmetrically eroded using the 3D thinning algorithm of the Skeletonize 3D plugin in ImageJ. The resulting 3D skeletons are evaluated using ImageJ's *AnalyzeSkeleton (2D/3D) plugin* to assess the number of branches, nodes, and end-points, and the order of nodes as described in sections 5-6 of the **ESI**. To determine the proportions of node orders, the inter-trabecular angle calculation program by Reznikov et al.[28] is used. We normalise X and C by dividing them by the analysed microstructure volume, V, which gives the corresponding densities χ and c.

For unconnected, cylindrical pores in one dimension, we obtain zero connectivity density c and a positive value for χ (7.23 μm$^{-3}$), which corresponds to n, the number of pores per unit volume V, since χ = n - c. For a single, connected pore network, n is given by N/V (N = 1 and V = 125 μm$^3$ yielding n = 0.008 μm$^{-3}$); therefore, χ and c are almost identical in magnitude but of opposite sign. For interconnected pores in two directions with N = 1, χ becomes negative (χ = -102.71 μm$^{-3}$, X = -12838.33) and the connectivity density, c, becomes positive (c = 102.71 μm$^{-3}$, C = 12839.33), indicating that the number density of redundant connections in the pore network has increased.

The PE microstructure is more connected (**Table II**; c = 143.16 μm$^{-3}$, C = 17894.67) than the reference microstructure with pores in 3 directions (c = 117.37 μm$^{-3}$, C = 14671.00), which has comparable geometrical pore radius $D_{50}$. In contrast, the PP separator exhibits a relatively small negative χ (χ = -7.43 μm$^{-3}$, X = -929.00) and a low connectivity (c = 7.44 μm$^{-3}$, C = 930.00), which can be understood by noting its straight pores with few redundant connections.

To further understand these trends in connectivity, we systematically analyse the proportions of node order, node and branch densities, percentage of end point branches, and average branch length (**Table III**). The reference separator pore networks with pores in two and three dimensions have a similar fraction of nodes of order 3-6 (section 6 in the **ESI**) and, as designed, zero end point branches within the volume. As larger numbers of perpendicular pore channels are introduced, the node and branch densities increase and the average branch length decreases. For the reference datasets, the pore dead-ends all are at the dataset's boundaries, whereas for the measured datasets, the pore dead-ends also appear within the volume.

**Table II**. Average values and standard deviations of porosity, topological invariant density, χ, and connectivity density, c, for the artificially generated microstructures (1D, 2D and 3D) and the imaged Targray PE16A (PE) and Celgard® PP1615 (PP) separator microstructures of edge lengths 5 μm each. The values for χ and c are calculated via the Minkowski functional, $M_3$.

| Parameter | 1D | 2D | 3D | PE | PP |
|---|---|---|---|---|---|
| Porosity [%] | 39.95±0.00 | 40.48±0.07 | 41.04±0.04 | 40.53±0.77 | 40.19±0.42 |
| Topological invariant density χ [μm$^{-3}$] | 7.23±0.00 | -102.71±0.16 | -117.36±1.15 | -143.15±6.88 | -7.43±0.51 |
| Connectivity density c [μm$^{-3}$] | 0.00±0.00 | 102.71±0.16 | 117.37±1.15 | 143.16±6.88 | 7.44±0.51 |

Consistent with its low connectivity, PP exhibits a lower node density and a larger average branch length than the PE separator. Analysis of the pore orientations (section 7 in the **ESI**) indicates that PP contains straight pores, while pores in the PE separator are also angled relative to one another. This difference in how pores are connected in PE and PP separators is further revealed by the different fractions of node orders. PE and PP exhibit ratios of 81:15 and 90:9, respectively for nodes of orders 3 and 4. The larger number of higher order nodes combined with the larger connectivity in PE compared to PP separator reflects the high redundancy of connections between nodes and higher spreading power. Thus, on a device level, transport through the separator pore network remains unchanged even if some pores are blocked. Finally, the PP separator exhibits a larger percentage of end point branches within the volume (31.55 %) than the PE separator (9.07 %), see section 8 in the **ESI**.

To understand the impact of these structural differences on battery performance, we perform steady-state Fickian diffusion simulations on the artificial and measured separator structures. We use cubic datasets of 3 µm edge length and iteratively calculate the solution of the Poisson equation on the electrolyte domain of the input geometry along the TP direction. We use Dirichlet boundary conditions at the end planes orthogonal to the TP direction and zero flux Neumann boundary conditions on all other boundaries and on the separator surfaces.[29,30] Inlet and outlet concentrations of 1.25 and 1.20 M are chosen based on the COMSOL simulation of C-rate dependence of electrolyte salt concentration for Li⁰|separator|LTO cells with Targray PE16A separator shown in the **ESI**.

If only the TP effective transport coefficient $\delta_{TP}$ calculated across a volume were considered, similar results would be would be obtained for the PE and PP separators as such calculations result in an overall value without local resolution. The effects of local variations in separator microstructure and disturbances (e.g., defects) cannot be resolved.

However, **Figure 4** shows the concentration profiles and density maps of these simulated, steady-state concentration gradients at different depths in the separator structure, which reveal the influence of separator topology on the concentration distributions. For 1D pores in the TP direction, all cylindrical pore channels have the same concentration at a given depth, so the concentration profile is a straight line. Upon adding more pores in a second and third direction, the concentration profile at a given depth broadens slightly (~2 mM and ~3 mM, respectively). For the PE separator, the concentration profile broadens to ~13 mM, indicating a variety of ion concentrations in different pores at a given depth. The concentration profile for the PP separator shows a broad distribution of concentrations at each depth. As marked by arrows, there are also regions where the same electrolyte concentration is found over close to 1 µm in length. This comes from dead-end pores, which extend in the TP direction but lack a connection with other pore channels.[31] Compared to the PE dataset, the PP dataset shows many more such threads, consistent with the network analysis (**Table III**).

Assuming PE separator thicknesses of 12-16 µm, we estimate a broadening of 45-60 mM. This corresponds to a range of 23-24% of the calculated concentration differences (~190-250 mM) at each depth, as outlined in section 9 in the **ESI**. In contrast, for the PP separator, a broadening corresponding to ~50 % of the concentration differences is found. This is consistent with the higher connectivity in the PE separator than the PP separator.

Regions of different electrolyte concentrations throughout the separator and impinging on the electrode may contribute to uneven lithium insertion in the electrode material, resulting in uneven expansion, diffusion induced stress and cracking,[13,32,33] as well as local overcharging or deep discharging. This can diminish battery performance and shorten battery life-time.

Therefore, we expect that a highly-connected structure reduces degradation in a battery (**Table II**).

This finding highlights the importance of knowing connectivity in a separator structure: the broadening of the local ion concentration distribution across a separator in the TP direction cannot be deduced from the effective transport coefficient $\delta_{TP}$.

**Table III**. Pore network properties for the artificially generated microstructures with pores in 2 and 3 directions, as well as for Targray PE16A (PE), and Celgard® PP1615 (PP). The values are averaged for three datasets of edge length 5 µm.

| Parameter | | 2D | 3D | PE | PP |
|---|---|---|---|---|---|
| **Proportion [%] of nodes of order** | 3 | 74.44±1.07 | 74.37±0.50 | 80.97±0.35 | 89.60±0.17 |
| | 4 | 24.00±0.26 | 22.07±0.53 | 15.16±0.14 | 9.28±0.16 |
| | 5 | 0.68±0.10 | 2.82±0.07 | 3.08±0.13 | 1.00±0.03 |
| | 6 | 0.01±0.01 | 0.57±0.03 | 0.60±0.05 | 0.10±0.04 |
| **Node density [µm⁻³]** | | 171.56±1.40 | 187.24±1.66 | 282.68±8.64 | 36.50±1.23 |
| **Branch density [µm⁻³]** | | 284.35±2.12 | 314.46±2.18 | 490.03±14.62 | 69.87±2.06 |
| **End point branches [%]** | | 0 | 0 | 9.07±0.39 | 31.55 ±0.81 |
| **Average branch length [nm]** | | 158.74±1.35 | 148.12±0.62 | 129.85±0.42 | 189.88±0.92 |

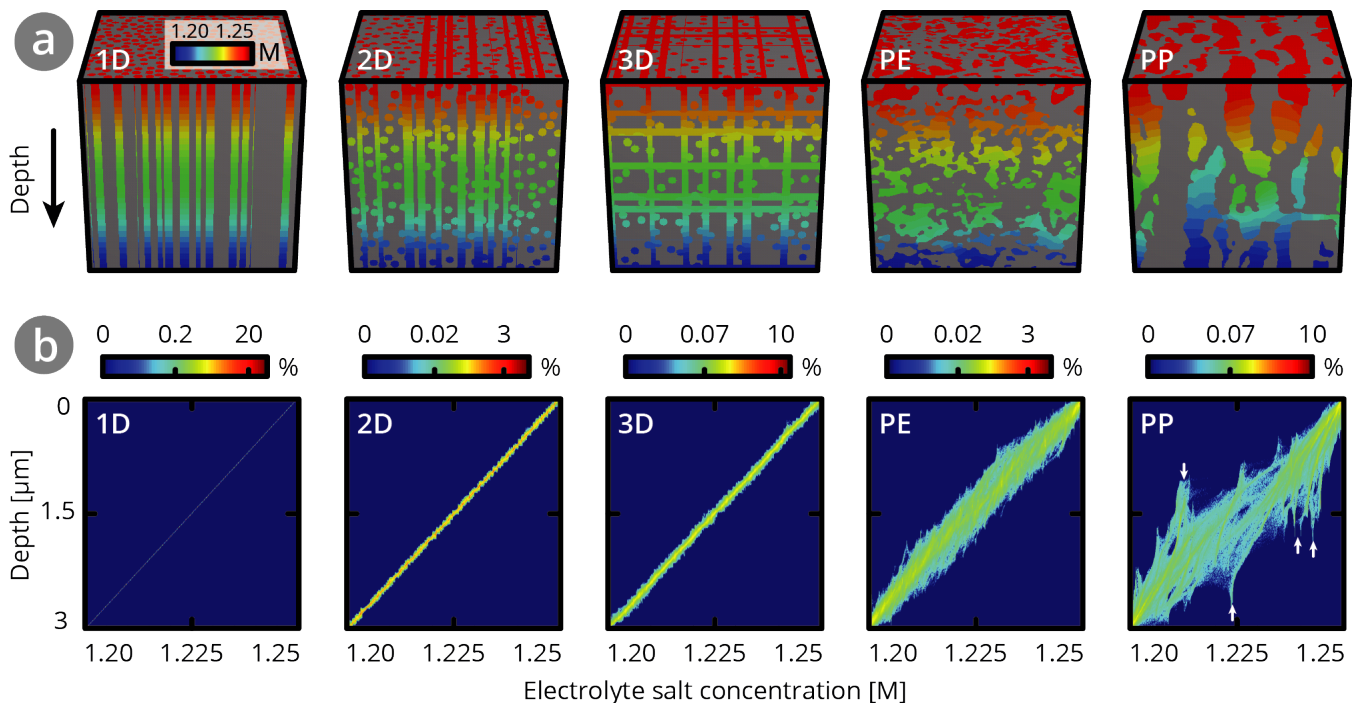

**Figure 4.** (a) Concentration profiles and (b) concentration density maps from steady-state Fickian diffusion simulations across the through-plane direction of artificially generated and recorded datasets of 3 μm edge length with a concentration difference of ~50 mM between top (1.25 M) and bottom (1.20 M).

In a next step, we assess how efficient the different separator topologies are at smoothing out in-plane ion concentration gradients that impinge on separator structures and are caused, e.g., by blocked pores. In a commercial lithium ion battery (schematic shown in **Figure 5a**), active particles in an electrode are typically on the order of 1 to 40 μm in diameter. This means that the electrode pore space structure has typical features approximately one to two orders of magnitude larger than the pore space of the separator. Direct contact between electrodes and separators can result in different concentrations of lithium ions in regions where the separator's pores are blocked by the electrode particles, and concentrations in regions where the electrode's and separator's pores meet. An example of an interface between a graphite electrode and a separator is shown in **Figure 5b**. Alternatively, a defect during separator manufacturing or battery assembly (e.g., agglomeration or contamination) as shown in **Figure 5c** may result in blocked areas and in decreased performance.[34] Pore-blocking defects can create ion-insulated regions, which locally may lead to high Li-ion concentrations and over-potentials at the distant separator interface. Local defects in separators lead to non-uniform charging and plating around the defect.[12]

To simulate these types of scenarios and gain an understanding of how high connectivity in a structure can help compensate for concentration gradients, we assume that a 3 μm circular object (electrode particle or defect) locally prevents electrolyte from impinging on the separator. **Figure 5e** shows the concentration density maps as in **Figure 4**, and **Figure 5f** shows the ion concentration profile at 1.5 μm depth for each sub-volume.

For structures with zero connectivity, the blocked pores do not contribute to the effective transport and form an ion-insulated region. For structures with intermediate connectivity in the range of >0 to ~100 μm$^{-3}$ (i.e., the 2D artificial structure or the PP structure, see sections 10-11 in the **ESI**), the pore network can compensate for the presence of defect structures at a depth of ~3 μm, while for microstructures with high connectivity in the range of 100 to 150 μm$^{-3}$ (i.e., the 3D artificial structure and PE) are only mildly affected by the presence of defect structures, and the fan-like distortion ends at around 1.5 μm. Due to its many redundant connections and slanted pores, PE is better in equalising in-plane ion gradients than the 3D artificial structures.

## Conclusions

In summary, we quantified the difference in the topological parameters and node structure of PE and PP separators of comparable porosity, TP tortuosity, and effective transport.

High connectivity of the pores, as found in PE separators, enables ion gradients present at the top of the separator to be smoothed out within a fraction of the separator thickness. A structure with multiple straight cylindrical channels, though offering excellent TP tortuosity and effective transport, has zero connectivity density and, due to the likely presence of defects, is more susceptible to Li plating if integrated into a lithium ion battery. In order to understand separator performance and optimise next generator separators for superior performance in cells, connectivity should be considered.

Beyond their function in describing homogenisation of ion concentration gradients through separators as described in detail here, topological and network-based analysis can also be used to predict how a structure will respond to mechanical or thermal stress[35]. By leveraging known trends in how a structure of a given topology shrinks under thermal stress[36], deforms in response to compressive or tensile stresses[37,38], or maintains connectivity despite closing of branches or nodes[35], it will be possible to predict a separator's response to many of the dynamic processes experienced during cell manufacturing and operation[39].

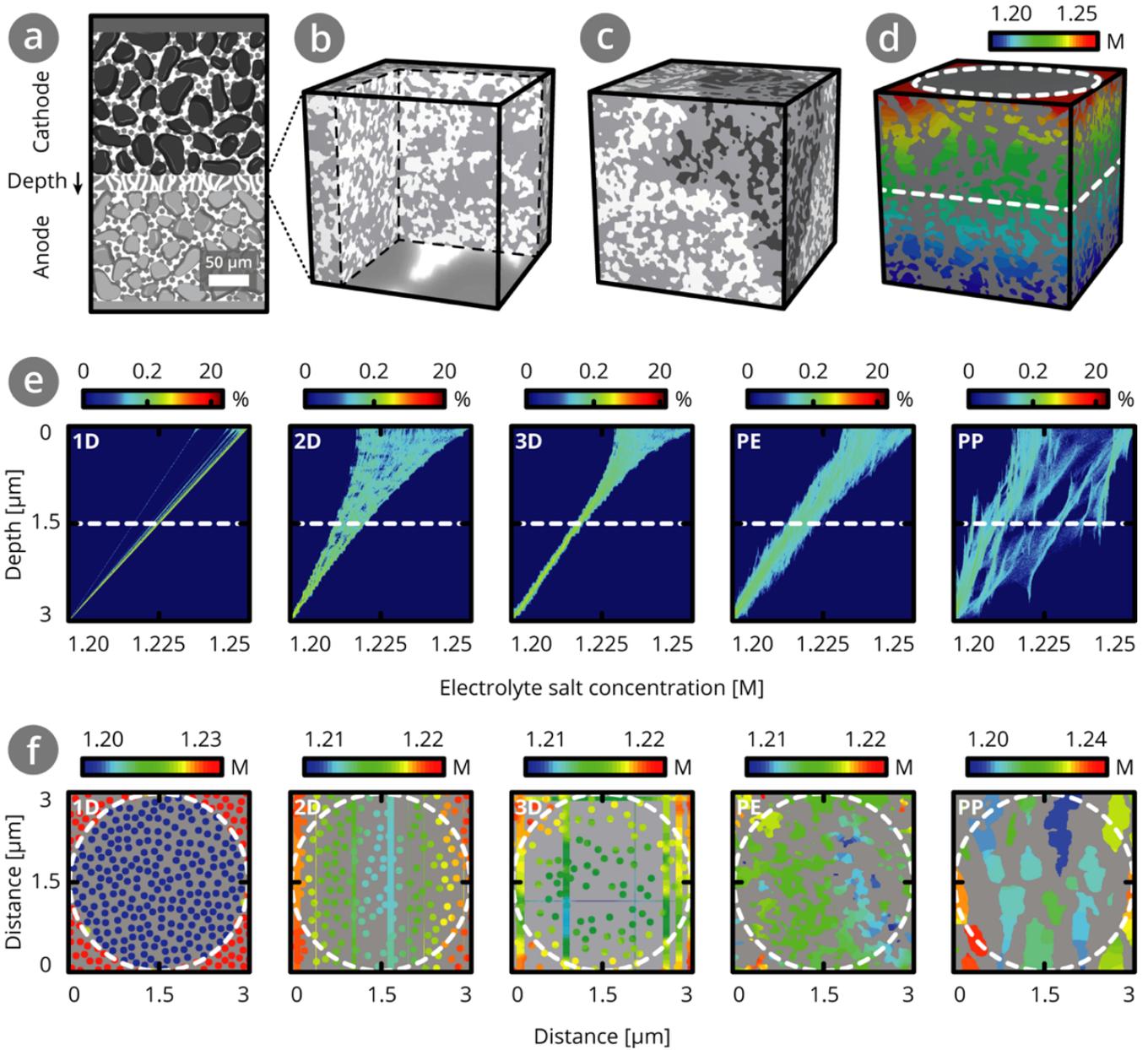

**Figure 5.** (a) Schematic of LIB setup with electrode particles touching the separator. (b) Rendering of separator volume of 3 µm edge length with separator-electrode interface. The electrode particles are of diameters 3-6 µm (i.e., up to three orders of magnitude larger than separator pores)[40] and can thus block a significant number of separator pores. (c) Rendering of separator volume of 3 µm edge length with blocked pores. (d) Concentration distribution from diffusion simulation across separator volume of 3 µm edge length with defect of diameter 3 µm on top (dashed circle) and at half depth (dashed line at 1.5 µm). (e) Concentration density maps for steady-state Fickian diffusion simulations across the through-plane directions of the reference datasets with pores in 1D, 2D, and 3D, and the imaged datasets of PE and PP (3 µm edge lengths). A circular defect structure of 3 µm diameter was placed on top of the structures and a concentration difference of ~50 mM was applied between top and bottom. (f) Ion concentration profiles at 1.5 µm depth (dashed white lines in **Figures 5d** and **e**) across sub-volumes of 3 µm edge length. The dashed circle represents the defect structure at 0 µm depth (see **Figure 5d**).

Connectivity density can be calculated via a Minkowski functional and is important when describing homogenisation of ion concentration gradients across microstructures. Similarly, other morphological and topological parameters are helpful when assessing surface interactions and effects[31,40,41]. Among such parameters are other Minkowski functionals, which correspond to a microstructure's surface area and curvature (described in detail in sections 2-3 in the **ESI**).

Finally, lithium ion battery separators are just one example of a component in energy and environmental systems that can benefit from the topological and network analysis presented here[42,43]. Connectivity can also improve understanding and design of separators in other electrochemical systems such as fuel cells[44,45] or ion-selective membranes for desalination[46,47], providing insights such as how thick a membrane should be or how transport paths can be designed to prevent mixing of product/reactant streams. Furthermore, beyond separator technology, we propose that all electrochemical systems (catalysis stacks for fuel generation, sensitised solar cells, lithium ion battery anodes and cathodes, etc.) can be viewed as interwoven electronic and ionic networks. An ideal system will have balanced networks at all length scales that bring together (or carry away) electrons and ions (or reactants and products) at equal rates while maintaining mechanical stability.

## Conflicts of interest
There are no conflicts to declare.

# Electronic Supporting Information

# Topological and Network Analysis of Lithium Ion Battery Components: The Importance of Pore Space Connectivity for Cell Operation

Marie Francine Lagadec, Raphael Zahn, Simon Müller, and Vanessa Wood*

Department of Information Technology and Electrical Engineering, ETH Zurich, Zurich CH-8092, Switzerland

* To whom correspondence should be addressed.
E-mail: vwood@ethz.ch

## 1. Effective transport properties of the Celgard® PP1615 separator

The Celgard® PP1615 separator was analysed as described in our previous work[1]. This separator has relatively large pore channels leading to a large representative volume element (RVE) with an edge length of ~3 µm (see **Figure S1a**). For the Targray PE16A separator dataset[2], the RVE edge length is ~2 µm as determined in our previous work[1]. For the PP RVE edge length of 3 µm, we determine a porosity ε of 40.19±1.03 % (**Figure S1b**), tortuosities $\tau_{TP}$ = 2.04±0.19, $\tau_{IP1}$ = 2.31±0.24, and $\tau_{IP2}$ = 24.89±6.15 (**Figure S1c**). This is also reflected in the effective transport coefficients $\delta_{TP}$ = 19.9±2.0 %, $\delta_{IP1}$ 17.6±2.1 %, and $\delta_{IP2}$ = 1.7±0.4 % (**Figure S1d**); effective transport in the IP2 direction is thus approximately ten times worse than in the TP or IP1 directions.

The pore networks of both the Targray PE16A and the Celgard® PP1615 separators consist of a single interconnected pore network. Small areas of non-connected pore space that might be present in these separators do not contribute to ionic transport across the separator and should be omitted for performance evaluations. Our imaging process relies on infilling the connected pore structure of the separator with material of a high imaging contrast[1]. This infilling process omits the non-connected pore space yielding a single interconnected pore network.

## 2. Extensive and intensive Minkowski functionals

The pore space of a binary structure is defined as P, which has an embedding space Ω (P ⊂ Ω, with Ω occupying the total dataset volume, $V_\Omega$). The pore space's boundary is δP, and its surface element for cylindrically-shaped structures is ds = R dz dφ. The *extensive* Minkowski functionals, $M_x(P)$ (with x ∈ {0, …, d} and d being the dimensionality of the structure of interest, here: d = 3), can also be expressed as *intensive* parameters (i.e., normalised functionals), $m_x(P)$.[4]

$$m_x(P) = \frac{M_x(P \cap \Omega)}{V_\Omega}$$

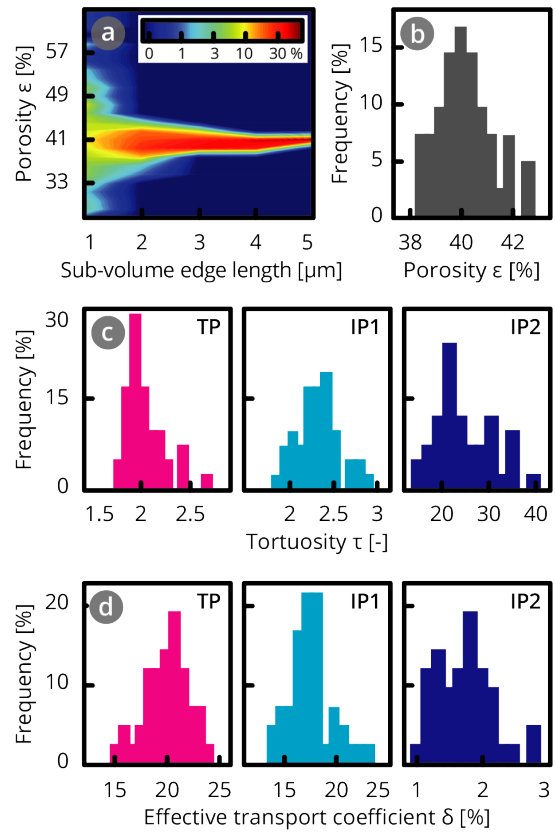

**Figure S1.** (a) Scale space analysis of the Celgard® PP1615 dataset[3] showing the convergence of the porosity distributions at different sub-volume sizes towards the mean value. (b) Porosity ε, (c) tortuosity τ, and (d) effective transport coefficient δ histograms for TP, IP1, and IP2 directions of the 3 µm edge length sub-volumes.

The first Minkowski functional, $M_0(P)$, corresponds to the pore volume, $V_{Pore}$:
$$M_0(P) = V_{Pore}(P).$$
The first normalised Minkowski functional, $m_0(P)$, corresponds to a structure's porosity, $\varepsilon$. For N non-intersecting cylindrical pores of height h and radius R (see **Figure 2a** in the main text), $M_0(P)$ becomes $N \cdot \pi R^2 \cdot h$.

The second Minkowski functional, $M_1(P)$, measures the interfacial area between pores and solid phase. The corresponding normalised functional is the specific surface area, $m_1(P)$.
$$M_1(P) = \int_{\delta P} ds$$
For N non-intersecting cylindrical pores of height h and radius R (**Figure 2a**), the integral yields $N \cdot 2 \pi R \cdot (R + h)$.

The third Minkowski functional, $M_2(P)$, measures the mean curvature, $H(P)$, over the interface and is a one-dimensional shape factor for 3D shapes. The corresponding normalised functional is called mean breadth density, $m_2(P)$.
$$M_2(P) = \frac{1}{2\pi} \int_{\delta P} H(P)\, ds = \frac{1}{4\pi} \int_{\delta P} \left[ \frac{1}{r_{min}} + \frac{1}{r_{max}} \right] ds$$
For N non-intersecting cylindrical pores of height h and radius R (**Figures 2a** and **3a**) the integral yields $N/2 \cdot (h + \pi \cdot R)$. Thus, for long, small pores, $M_2(P)$ becomes $N \cdot h/2$. For complex structures with interconnectivity in several directions (**Figures 3b-c**), $M_2(P)$ becomes more complex, and can be calculated numerically, but no longer analytically.

The fourth Minkowski functional, $M_3(P)$, measures the Gaussian curvature, $K(P)$, over the interface (thus, the total curvature), and is proportional to the Euler-Poincaré characteristic, $X(P)$, a topological invariant.
$$M_3(P) = \int_{\delta P} K(P)\, ds = \int_{\delta P} \left[ \frac{1}{r_{min} \cdot r_{max}} \right] ds = \\ = 4\pi \cdot X$$
In 3D, the characteristic X is linked to the Betti numbers $\beta_0$ (number of objects N), $\beta_1$ (connectivity C) and $\beta_2$ (number of enclosed cavities), and, for a voxel-based dataset, to $N_V$ (number of vertices), $N_E$ (number of edges), $N_F$ (number of faces), and $N_{vox}$ (number of voxels, or solids).[4,5]
$$X = \beta_0 - \beta_1 + \beta_2 = N_V - N_E + N_F - N_{vox}$$
For percolating networks of pores and solid (i.e., without enclosed cavities), the characteristic X of the pore space can simply be expressed by the number of pores, N ($N \geqq 0$), and the connectivity, C ($C \geqq 0$).
$$X = N - C$$
The corresponding normalised parameters, $\chi$ and c, are the Euler-Poincaré characteristic density and the connectivity density.

### 3. Minkowski functionals of separator microstructures

Minkowski functionals have previously been linked to transport related parameters. The shape factor, in combination with the surface area, provides a first approximation of the diffusion coefficient[6]. Structures with high connectivity have large node and branch densities, and higher order nodes are associated with more spreading power.[7] Meanwhile, in diffusion simulations, branches that dead-end do not contribute to effective transport through the structure.[8]

The intensive Minkowski functionals (i.e., normalised to the analysed volume and designated here with $m_x$) are listed in **Table T1**. As designed, the reference separator microstructures replicate porosities well within the specified porosity of 40±2 and 40±5 % of the PE and PP separators. The specific surface area of the reference datasets slightly decreases as more pores are added in a second and third perpendicular direction. The specific surface area ($m_1$) of the PE separator (11.72 µm$^{-1}$) is almost identical to $m_1$ of artificially generated microstructures with pores in three perpendicular directions (11.62 µm$^{-1}$). This is expected since pore size as well as porosity were chosen to match the parameters of the PE separator. It shows that cylindrical pore segments are a good approximation for the pore shape in the PE separator. For the PP separator, the value for $m_1$ is lower (5.22 µm$^{-1}$) due to the larger pores in PP.

**Table T1**. Average values and standard deviations of the intensive Minkowski functionals $m_0$, $m_1$, $m_2$, as well as Euler-Poincaré characteristics and connectivity densities χ and c, respectively, for the artificially generated microstructures (1D, 2D and 3D) and the imaged Targray PE16A (PE) and Celgard® PP1615 (PP) separator microstructures of edge lengths 5 µm each. The values for χ and c are calculated via the Minkowski functional $M_3$.

| Parameter | 1D | 2D | 3D | PE | PP |
|---|---|---|---|---|---|
| Porosity $m_0$ [%] | 39.95±0.00 | 40.48±0.07 | 41.04±0.04 | 40.53±0.77 | 40.19±0.42 |
| Specific surface area $m_1$ [µm$^{-1}$] | 13.94±0.00 | 12.11±0.02 | 11.62±0.02 | 11.72±0.13 | 5.22±0.14 |
| Shape factor density $m_2$ [µm$^{-2}$] | 17.40±0.01 | 7.82±0.03 | 5.44±0.03 | 6.68±0.35 | 1.54±0.07 |
| Topological invariant density χ [µm$^{-3}$] | 7.23±0.00 | -102.71±0.16 | -117.36±1.15 | -143.15±6.88 | -7.43±0.51 |
| Connectivity density c [µm$^{-3}$] | 0.00±0.00 | 102.71±0.16 | 117.37±1.15 | 143.16±6.88 | 7.44±0.51 |

$m_2$ can be interpreted as a measure of shape of the surface. For the all structures (reference and real), the values are positive, indicating that the shape of the pore surface is on average convex.[9,10] For the reference datasets with cylindrical pores in one direction, the shape factor scales with the number of pores $N$ and with $h+\pi \cdot R$, where $h$ is the pore length and $R$ is the pore radius. As more pores are introduced in a second and third perpendicular direction, the shape factor density decreases (from 17.40 µm$^{-2}$ to 7.82 µm$^{-2}$ in 2D and 5.44 µm$^{-2}$ in 3D).

The PE separator's shape factor density (6.68 µm$^{-2}$) is comparable to the ones of the datasets with straight pores in 2 and 3 directions, while the PP separator's shape factor density is lower (1.54 µm$^{-2}$) because of its higher proportion of concave regions at the pore surface.

### 4. Algorithm to generate artificial separator geometries

The following paragraph describes an algorithm to generate an artificial separator geometry consisting of solid elements and pore space. The pores are cylindrical and their orientation is always parallel to the IP1, IP2, or TP direction. At first, a cuboid with desired dimensions is defined. In three different datasets, we create pores in the TP direction (1D pore directionality), pores parallel to the TP and IP1 direction (2D pore directionality), and pores parallel to the TP, IP1, and IP2 axis (3D pore directionality). The desired porosity of 40 % is divided by the number of pore directions (one, two, or three) in order to get the same porosity in all directions. The pore generating process consists of two major steps.
(i) On the face orthogonal to each desired pore direction, circles with a predefined radius (130 nm) are generated in an iterative manner. The number of circles is set by the porosity. The location of the circles is random with the only constraint that they cannot touch or intersect.
(ii) As soon as the necessary quantity of circles is created, they are extended through the entire separator producing the pores.
If more than one pore direction is wanted, the total porosity might be smaller than the addition of the directional porosities since pores may intersect. In this case, a new pore generating iteration is induced (starting from (i)) whereby the shortage in porosity is converted into the new number of circles to be created. This procedure runs until the target porosity of 40±2 % is met.

While the artificially generated 1D microstructures consist of individual pores that are not interconnected amongst each other, the artificially generated 2D and 3D microstructures are strongly interconnected and form a single connected pore network. This interconnectivity is not an implicit result of our algorithm for creating artificial microstructures, however, at 40 % porosity, it is extremely unlikely for cylindrical pores to penetrate 5 µm (= 500 voxels) thick structures without crossing another pore (probability ~ $(0.6)^{500}$). Therefore, none of the used artificially generated 2D and 3D microstructures contain isolated, non-interconnected pores.

Artificially generated 2D and 3D microstructures contain high fractions of third order nodes (see **Table III**). Nodes of order four (or higher) are only created if the central skeleton lines of two (or more) pores intersect in one single point. For pores of final diameter, such events have a low likelihood, and for most intersecting pores, the central skeleton lines of these pores will not intersect. Thus, several third order nodes are created instead of one single higher order node.

### 5. Shape analysis

For a 2D network, the connectivity C (number of loops) can be calculated via
$$C = N_B - N_N - N_{EP} + 1$$
with $N_B$ being the number of branches, $N_N$ being the number of nodes, and $N_{EP}$ being the number of end points.[11] The description for connectivity using branches, end-points, and nodes is valid only in 2D; in 3D, the correct description uses edges, faces, and vertices of the single voxels. For our structures with 1D and 2D pore directionality, the values for connectivity using the descriptions for 2D and 3D are the same. For the artificially generated pore structures in 3D and the recorded datasets, the values obtained via the 2D description are off by less than ±1 % compared to the values obtained via the 3D description.

## 6. Network analysis log-log-plots

In network theory, it is common to assess the node order distributions on a log-log plot. A Poisson distribution indicates a random network and power law distribution indicates a scale-free network.

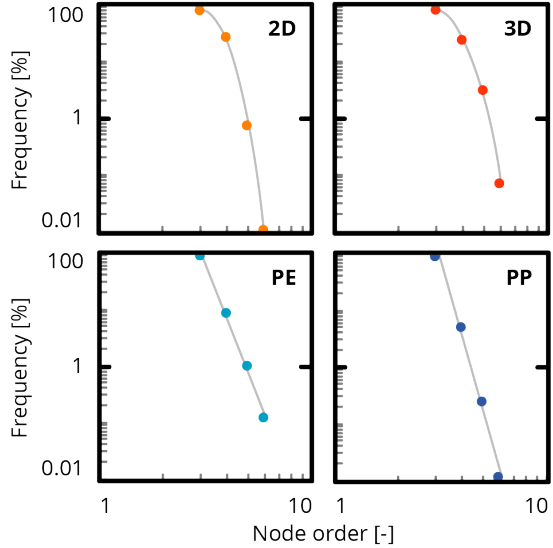

**Figure S2.** Log-log-plots of the node order distribution of the 2D and 3D reference datasets and the imaged PE and PP datasets.

**Figure S2** shows that node order distribution in the 2D and 3D dataset follows an exponential distribution, while the distributions of the PE and PP datasets seem to follow a power law. However, to quantify the scaling of the node order, the distribution should exhibit a linear relationship on a log-log-plot over at least two orders of magnitude in both the x and y axes.[12] In standard processing of voxel-based data only nodes of order 3-6 can be reliably identified.[13] The x axis of **Figure S2**, therefore, spans less than one order of magnitude.

## 7. Pore orientation analysis

We determine the pore orientation angle distribution using ImageJ's *Directionality plugin* for the non-processed datasets for the PE (**Figure S3a**) and PP (**Figure S3b**) separators. As illustrated in the left-most images, the orientation angle is calculated for all pores in a plane, and each plane is indexed by the slice number in a specific direction (TP, IP1, or IP2). A vertical cut through any of the pore orientation angle distribution density plots (three plots to the right), would yield a histogram that represents the pore orientation distribution for that slice. For a slice (i.e., plane) with pores perpendicular to the slice, the orientation angle is 0°.

In the PE separator, for slices along the TP direction, the peak of the pore orientation distribution varies between 0° and 180°. This may be attributed to the presence of fibres in the IP direction located at different separator depths. For slices in the IP1 and IP2 directions, the pore orientation distributions are broadly and asymmetrically centred above 0°, indicating pores slanted at many different angles.

For PP, slices along the TP and IP1 are similar with orientation angle histograms centred around 90° due to its straight pore in the TP and IP1 directions.

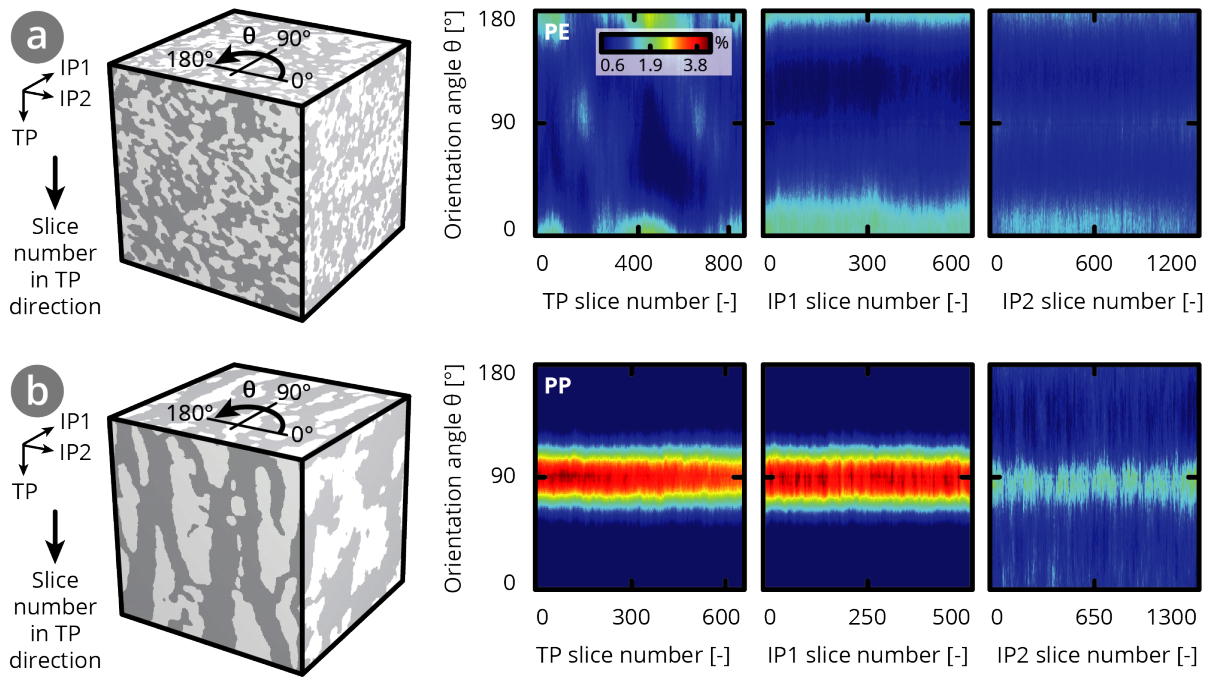

**Figure S3.** Orientation angle distributions across TP, IP1, and IP2 directions of the greyscale value, non-processed datasets of (a) Targray PE16A, and (b) Celgard® PP1615 separators.

## 8. End point analysis

The skeletonized pore space of the reference and imaged datasets of 5 μm edge length gives the total number of branches in the sub-volume, $N_B$. To obtain the number of network branches, $N_n$, we prune the skeletonized datasets using ImageJ's *AnalyseSkeleton 2D/3D plugin*. Subtracting the pruned skeleton from the original skeleton gives the number, $N_{EP*} = N_B - N_n$, of the end point branches within the volume and their coordinates. To account for end points that stem from cropping the datasets, we discard end point branches with coordinates within 5 voxels of the sub-volume's surface. We determine the end point density as the number of end point branches per volume ($N_{EP*} / V$), and the percentage of end point branches as the fraction of end point branches and the total number of branches ($N_{EP*} / N_B$).

## 9. Steady-state diffusion simulations

We simulate the C-rate dependence of the electrolyte salt concentration gradient across a Li$^0$|separator|LTO cell as shown in **Figure S4** and as described in our earlier work.[14] At 1C, a concentration difference of 0.25 M builds up across the 16 μm thick separator; this corresponds to a concentration difference of ~50 mM across a sub-volume of 3 μm edge length and to inlet and outlet concentrations of 1.25 and 1.20 M, respectively. We plot the broadening of the electrolyte salt concentration at each depth in the TP direction for artificially generated and imaged datasets (**Figure 4** in the main text).

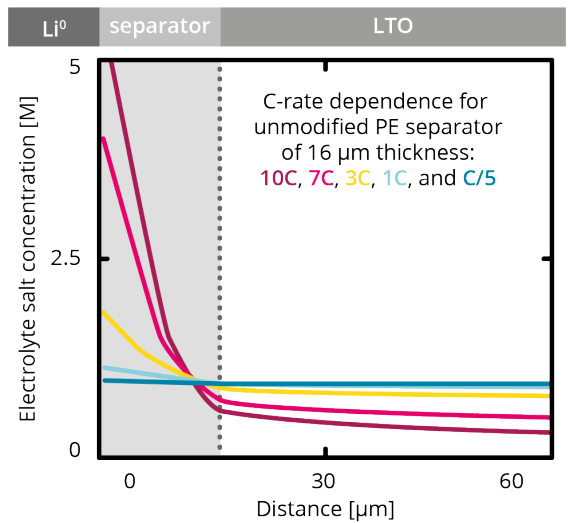

**Figure S4.** C-rate dependence of electrolyte salt concentration for Li$^0$|separator|LTO cells with Targray PE16A separator.

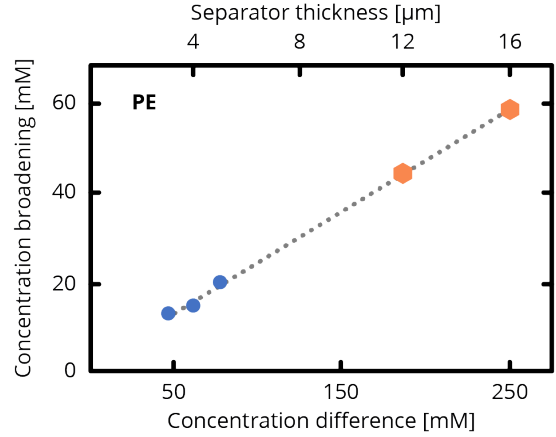

**Figure S5.** Calculated concentration broadening as function of sub-volume edge length of separator volume and corresponding concentration difference.

In **Figure S5**, we plot the concentration broadening (i.e., histogram width at half-depth of sub-volume) calculated for sub-volumes of 3, 4 and 5 μm (blue data points) against the concentration difference across a separator sub-volume (lower x-axis, range of values from **Figure S4**) and the edge length of the separator sub-volume (upper x-axis), and extrapolate the broadening of the electrolyte salt concentration to edge lengths of 12-16 μm (orange data points).

## 10. Adding nanofibers to Celgard® PP1615 separator geometries

Due to resolution limitations, the measured PP dataset does not feature the amorphous PP nanofibers spanning the large pore channels (partially visible in the SEM in **Figure 1b**). To assess how the presence of nanofibers in the large pore channels affects effective transport and topological properties of the separator geometry, cylindrical nanofibers with a diameter of 40 (thin nanofibers) and 60 nm (thick nanofibers) and distances of 50-70 nm are added to the pore channels of the recorded PP dataset. The diameter and the distances are estimated from FIB-SEM cross-sectional images.

At first, the existing separator geometry is loaded and rotated such that the direction along which the fibres have to be created corresponds to the IP2 axis. Then, the geometry is up-scaled isotropically in order to decrease voxel size followed by three-dimensional smoothing with a Gaussian kernel. Start points of a given number of fibres are randomly generated in the IP direction. The direction of each fibre is slightly deflected at random such that a direction distribution is created. A fibre ends as soon it (re-)enters the other side of the separator. All fibres are dilated to the desired radius and in a final step, the resulting structure is smoothed again.

**Table T2.** Normalised Minkowski functionals $m_0$, $m_1$, $m_2$, as well as Euler-Poincaré characteristic and connectivity densities, χ and c, respectively, for a PP separator microstructure of edge length 3 μm as imaged, and with thin and broad nanofibers. The values for χ and c are calculated via the Minkowski functional $M_3$.

| Datasets | PP1615 | PP1615 with nanofibers of diameter 40 nm | PP1615 with nanofibers of diameter 60 nm |
|---|---|---|---|
| Porosity $m_0$ [%] | 39.38 | 38.56 | 35.80 |
| Specific surface area $m_1$ [μm$^{-1}$] | 5.30 | 7.66 | 7.08 |
| Shape factor density $m_2$ [μm$^{-2}$] | 1.82 | -3.31 | -0.07 |
| Topological invariant density χ [μm$^{-3}$] | -6.70 | -94.07 | -54.89 |
| Connectivity density c [μm$^{-3}$] | 6.74 | 94.11 | 54.93 |

### 11. Celgard® PP1615 separator geometries without and with nanofibers

**Figure S6** shows that there is little difference between the calculated density plots for the electrolyte salt concentrations across a sub-volume of the PP1615 separator as imaged and with added nanofibers.

Thus, we conclude that – from a geometric perspective – the effective transport properties are not affected significantly by the presence of the PP nanofibers. The effect of the nanofibers cannot be neglected when modelling the mechanical properties of PP separators, as shown by Xu et al.[15] for Celgard® 2400 separator.

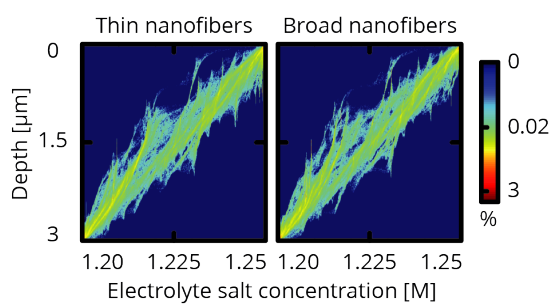

**Figure S6.** Electrolyte salt concentration across a sub-volume of 3 μm edge length of the Celgard® PP1615 separator dataset with artificially added thin (40 nm) and broad (60 nm) nanofibers.

The calculated Minkowski functional densities listed in **Table T2** show that adding nanofibers to the PP dataset results in a lower porosity, an increased specific surface area, a negative shape factor, and a more negative topological invariant. The latter corresponds to a more positive connectivity density (up to ~100 μm$^{-3}$), which is below the calculated connectivity density of PE. Since the surface integral of the mean curvature can be interpreted as the average of the mean curvature, a more positive shape factor indicates the presence of more convex parts. A negative shape factor like in the case of added nanofibers indicates thus more concave regions. For **Table T2**, we calculate the intensive Minkowski functionals for a single dataset of 3 μm edge length; in **Table II** in the main text, we calculate the average and standard deviation of the intensive Minkowski functionals for three datasets of 5 μm edge length.